\documentclass[a4paper]{easychair}

\usepackage{doc}
\usepackage{makeidx}
\usepackage{algpseudocode}
\usepackage{caption}
\usepackage{tikz}
\usetikzlibrary{arrows,positioning,shapes.misc}

\usepackage{algorithm}

\usepackage{amsmath,amssymb}
\newcommand{\packcopy}{\mathbf{1}}
\newcommand{\rec}[1]{\mathbf{rcfg(#1)}}

\newcommand{\mcal}[1]{\mathcal{#1}}

\newcommand{\step}[1]{\xrightarrow{#1}}

\makeatletter
\def\xtwoheadrightarrowfill@{%
 \arrowfill@\relbar\relbar{\rightarrow\mkern-15mu\rightarrow}}
\newcommand*\xtwoheadrightarrow[2][]{%
 \ext@arrow 0099\xtwoheadrightarrowfill@{#1}{#2}}
\newcommand{\steps}[1]{\xtwoheadrightarrow{#1}{}}

\newcommand{\drop}{\mathbf{0}}
\newcommand{\NetKATnoDup}{\textnormal{NetKAT}^{-{\bf{dup}}}}
\newcommand{\zeroq}{\bot}
\newcommand{\Seq}{\mathop{;}}
\newcommand{\Par}{\mathop{||}}

\begin{document}

%
\title{Counterfactual Causality in Networks 
\thanks{Abstract accepted at the 33rd Nordic Workshop on Programming Theory, NWPT 2022.}}



%
\author{
Georgiana Caltais \and Can Olmezoglu
}

\institute{
University of Twente, The Netherlands \\
\email{g.g.c.caltais@utwente.nl , c.olmezoglu@utwente.nl}\\
 }


\clearpage

\maketitle

%


%
%

\pagestyle{empty}

\section{Introduction \& Background}
\label{sect:introduction}
The main objective of an engineer is to build systems which follow a predefined behaviour. Explaining when a system fails to follow through that behaviour has thereby gained a lot of attention as engineering rose to prominence. In this abstract, we propose a framework for explaining violations of safety properties in Software Defined Networks (SDNs), using counterfactual causal reasoning~\cite{DBLP:journals/corr/abs-1301-2275}. 

SDN has gained a lot of traction due to its increased network management and programmability, achieved by decoupling the control plane from the data plane \cite{DBLP:journals/cacm/Kirkpatrick13}, in contrast with traditional networks. SDN technologies can play an important role in solving issues concerning big data applications, including data processing in cloud data centers, optimisations and data delivery. In this abstract, we focus on DyNetKAT --a rigorous framework for modelling and analysing (multi-)packet forwarding within an SDN, and communication between data and control planes-- introduced in~\cite{DBLP:conf/fossacs/CaltaisHMT22} by a subset of the authors.
DyNetKAT is based on NetKAT~\cite{DBLP:conf/popl/AndersonFGJKSW14}, a minimalist language based on Kleene Algebra with Tests, supported by a sound and complete axiomatisation. Packets in (Dy)NetKAT are encoded as sets of fields and associated values $\{f_1 = v_1,\, \ldots f_n = v_n\}$. NetKAT can model the forwarding of a single packet within a network, and includes constructs such as dropping of packets ($\drop$), acceptance of packets ($\packcopy{}$), multicast ($+$), packet fields modification ($f \leftarrow n$) and repeated application ($^*$) of these policies. In addition, NetKAT can be used to build packet histories using the {\bf{dup}} construct, which was dropped in~\cite{DBLP:conf/fossacs/CaltaisHMT22}. DyNetKAT extends NetKAT with channel-based communication ($\Par,\, x?p,\, x!p$) of flow tables ($p$) between the data and control planes (with synchronous communication of $p$ on channel $x$ encoded as $\rec{x,p}$), no-behaviour policies ($\zeroq$), non-deterministic choice ($\oplus$), recursive specifications ($X$) and multi-packet semantics in the context of a sequential composition operator ($\Seq$) that marks the fetching of a new packet in the packet queue. In contrast with NetKAT, DyNetKAT has an operational semantics that entails LTS models, and a sound and ground complete axiomatisation in the style of the Algebra of Communicating Systems (ACP). The syntax of DyNetKAT is: 
$
     \mathit{N} ::=  {\NetKATnoDup} ~~~
     \mathit{D}  ::= 
     { \zeroq } \mid
     \mathit{N} \Seq \mathit{D}
     \mid  x?\mathit{N} \Seq D \mid 
     x!\mathit{N} \Seq D \mid \mathit{D} \Par \mathit{D} \mid
     D \oplus D \mid X \textnormal{ with }  
      X \triangleq D
$. The complete framework is defined in~\cite{DBLP:conf/fossacs/CaltaisHMT22}.
Similarly to~\cite{DBLP:conf/fossacs/CaltaisHMT22}, we consider guarded DyNetKAT specifications that can be reduced to equivalent expressions in head normal form (Lemma 7 in~\cite{DBLP:conf/fossacs/CaltaisHMT22}). This, in turn, guarantees the existence of finite LTS models for DyNetKAT specifications with finite number of recursive variables, finite sets of channel names and packet fields over finite domains.

We base our SDN safety failure explanations on the so-called counterfactual causality, or actual causality, introduced in the seminal work~\cite{DBLP:journals/corr/abs-1301-2275}, and adapted in~\cite{DBLP:conf/tase/BonsangueCFT22} to the context of finite automata (as SDN models) and regular expressions (as a language for defining safety properties).
 Intuitively, (a sequence of) events $c$ are considered causal with respect to the realisation of a hazard $e$ whenever (i) $c$ is necessary for $e$ to happen, (ii) $c$ not happening entails $e$ not happening (this is known as the counterfactual test), (iii) there is no $c'$ ``simpler" than $c$ that can satisfy the conditions above. In addition, there might be the case that despite $c$ being observed, $e$ does not happen due to some other cancelling actions (e.g., the forest does not burn down, despite the lightning, because the firefighters arrive on time). Such situations are modelled by means of contingencies in~\cite{DBLP:journals/corr/abs-1301-2275} or events causal by their non-occurrence in~\cite{DBLP:conf/tase/BonsangueCFT22,caltais2020causal}. 
The causal analysis in~\cite{DBLP:conf/tase/BonsangueCFT22} is performed in the context of FA models, and safety violations, or hazards encoded as regular expressions defined in the standard fashion:
$e \,::=\,  0 \mid 1 \mid a \mid e \Seq e \mid e+e \mid e^*$. 
The computed causes are words, or decorated traces $w_0\,a_0\,w_1\,a_1 \ldots a_n\,w_n$ where $a_0\,a_1 \ldots a_n$ is a word which, if executed, leads to the hazard $e$, and $w_i$ ranges over contingencies that disable the hazard. Note that $w_i$ play an important role in describing fixes, or alternative safe scenarios.

\noindent
{\bf{Our contribution.}}
First, we devise and implement an algorithm that computes the LTS models of DyNetKAT programs. Then, we transform the aforementioned LTSs into FA models in a straightforward fashion, by handling every state as accepting. The generated FAs can be further analysed according to the causal inference machinery in~\cite{DBLP:conf/tase/BonsangueCFT22}. We explain our approach based on a running example --a faulty virtual circuit that allows illegal packet forwarding--. 

\vspace{-12pt}
\section {Running Example}
\label{sec:runningexample}
\begin{figure}[h]
\centering
\begin{minipage}{.5\textwidth}
  \centering
  \includegraphics[width=.4\linewidth]{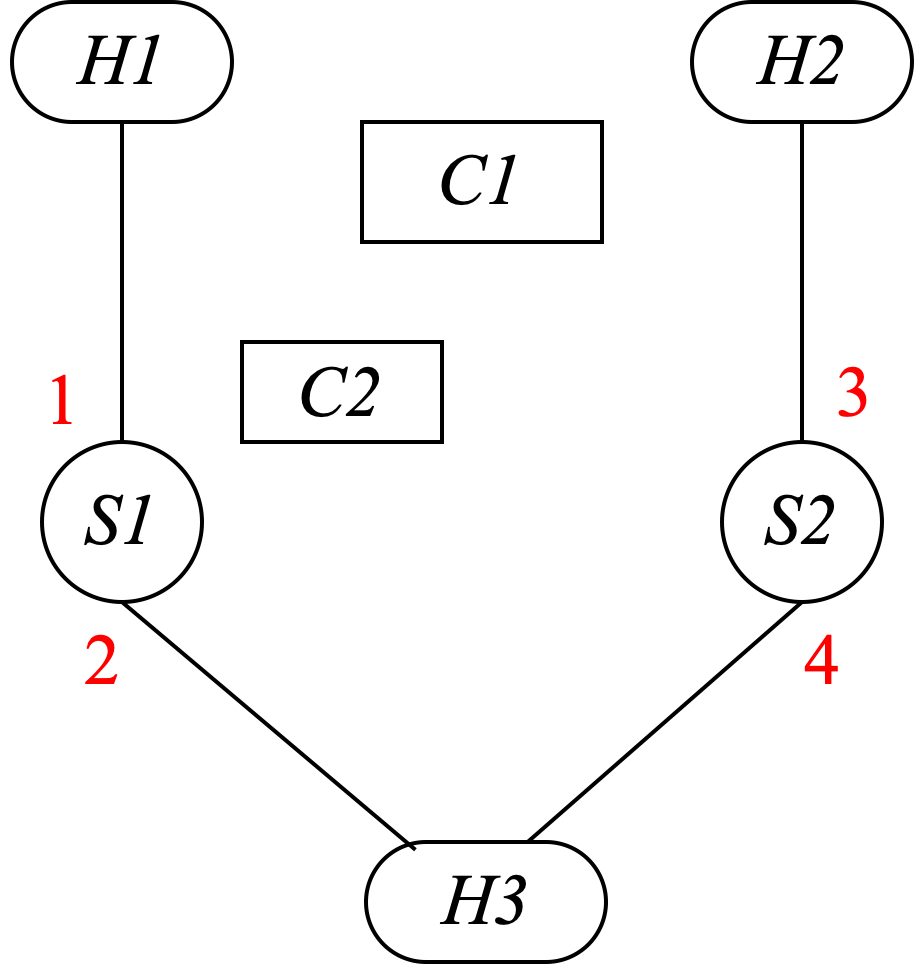}
  \captionof{figure}{The Virtual Circuit}
  \label{fig:vc_topology}
\end{minipage}%
\begin{minipage}{.5\textwidth}
\small
\begin{equation}
\label{eq:virtualcircuit}
\begin{aligned}
      & C1 \, \,  \triangleq NoVirtualCircuit!\textbf{1} ; C1 \oplus \\
      & \, \, \, \, \, \, \, \, \, \, \,  \, \, \, \,  \, \,  VirtualCircuitReq?\textbf{1};C1' \\
      & C1' \triangleq VirtualCircuitEnd?\textbf{1};C1 \\
      & C2 \, \,  \triangleq VirtualCircuitReq!\textbf{1} ; C2'  \\
      & C2' \triangleq VirtualCircuitEnd!\textbf{1};C2 \\
      & S1 \, \, \, \, \triangleq VirtualCircuitReq?\textbf{1};S1' \\
      & S1'\, \, \triangleq ((port=1) . (port \leftarrow  2));S1'\oplus \\
      & \, \, \, \, \, \, \, \, \, \, \, \, \, \, \,  \, \, VirtualCircuitEnd?\textbf{1} ;S1 \\
      & S2 \, \, \, \,\triangleq NoVirtualCircuit?\textbf{1};S2' \\
      & S2'\, \, \triangleq ((port=3) . (port \leftarrow 4));S2 \\
      & Init \, \triangleq C1 || S1 || S2 || C2
\end{aligned}
\end{equation}
\end{minipage}
\end{figure}

A virtual circuit is created for the delivery of a bit stream between a source host and a destination host \cite{DBLP:books/daglib/0010824}, denoted by $H1$ and, respectively, $H3$ in the example of Figure~\ref{fig:vc_topology}. However, when necessary, another host such as $H2$ in Figure~\ref{fig:vc_topology} should be able to send external packets to $H3$, provided that $H3$ is not currently receiving a bit stream. Controller $C1$ oversees the network and sends messages to network devices $C2$, $S1$ and $S2$, deciding when packets from $H2$ are forwarded or whether a virtual circuit between $H1$ and $H3$ can be initiated. For example, if $H2$ wants to send something to $H3$, the switch $S2$ connecting $H2$ and $H3$  checks whether there is a virtual circuit between $H1$ and $H3$ by querying $C1$. When a virtual circuit needs to be initiated, $C2$ informs $C1$ to make sure other packets are not being sent while the virtual circuit is active. After receiving this information, $C1$ stops allowing external packets from $H2$ to be forwarded to $H3$. 
In equation~(\ref{eq:virtualcircuit}) we provide a DyNetKAT formalism for the running example.
\par
A hazardous situation in the running example can happen when a virtual connection between $H1$ and $H3$ is active and processing a packet $\sigma_1$ into $\sigma_2$, indicating switch $S1$ forwarding the package from port $\textcolor{red}{1}$ to port $\textcolor{red}{2}$, depicted in red color in Figure \ref{fig:vc_topology}. As this forwarding operation is happening, a packet $\sigma_3$ from $H2$ is processed into $\sigma_4$ and forwarded to $H3$ by $S2$. As the circuit is already active and using most of the resources of $H3$, the arrival of $\sigma_4$ at $H3$ might lead to an overflow error.

\begin{figure*}
	 \resizebox{\columnwidth}{!}{
	\begin{tikzpicture}[scale=0.8,auto,node distance=2.1cm, dot/.style={}, node/.style={rectangle,draw,minimum height=1cm}]
	\node[node] (m0) {$n_0: (C2 \Par S2 \Par C1 \Par S1, \sigma_1 {::} \sigma_3 {::} \langle \rangle, \langle \rangle)$};
    \node[dot] (d1) [right=1cm of m0] {$\ldots$};
	\node[node] (m1) [right=1cm of m0] {$n_1: (C2 \Par S2' \Par C1 \Par S1, \sigma_1 {::} \sigma_3 {::} \langle \rangle, \langle \rangle)$};
	\node[node] (m2) [below=1.7cm of m1] {$n_2: (C2' \Par S2' \Par C1' \Par S1', \sigma_1 {::} \sigma_3 {::} \langle \rangle, \langle \rangle)$};
	\node[node] (m3) [right=1.7cm of m2] {$n_3: (C2' \Par S2' \Par C1' \Par S1', \sigma_3 {::} \langle \rangle, \sigma_2 {::} \langle \rangle)$};
	\node[node] (m4) [right=1.3cm of m1] {$n_4: (C2' \Par S2 \Par C1' \Par S1', \langle \rangle, \sigma_4 {::}\sigma_2 {::} \langle \rangle)$};
	\path[->]
	(m0) edge [bend left] node [pos=0.55,align=center] {$\small{NoVirtualCircuit?\packcopy{},}$ \\ $\small \rec{NoVirtualCircuit,\packcopy{}}$} (m1)
	(m1) edge [] node [align=center] {$\small {VirtualCircuitReq!\packcopy{},}$ \\
	$\small \rec{VirtualCircuitReq,\packcopy{}}$} (m2)
	(m2) edge [bend left] node [] {$\small \rec{VirtualCircuitEnd,\packcopy{}}$} (m1)
	(m2) edge [] node [align=left] {$\small(\sigma_1,\sigma_2)$} (m3)
	(m3) edge [] node [align=left] {$\small(\sigma_3,\sigma_4)$} (m4);
	\end{tikzpicture}    
	}
  \caption{Virtual Circuit LTS (excerpt) }
  
  \label{fig:stateful_lts}
\end{figure*}
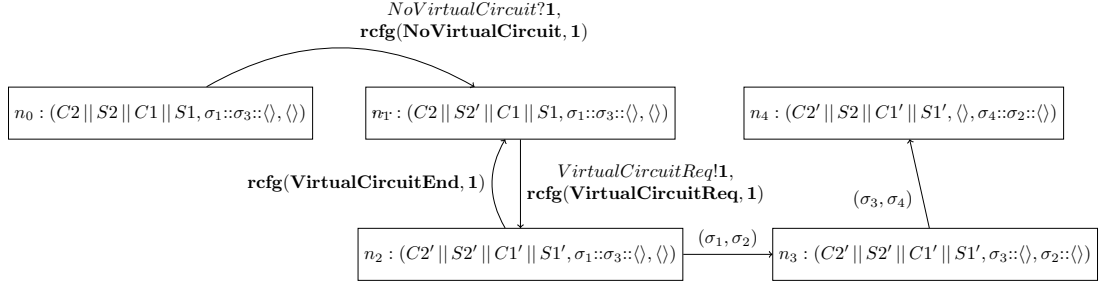

The LTS behavioural model of the DyNetKAT program in~Figure~\ref{fig:vc_topology} can be devised according to the DyNetKAT operational semantics. An excerpt of this LTS is provided in Figure~\ref{fig:stateful_lts}.
For instance, the trace $\rec{NoVirtualCircuit,\packcopy{}} \,\, \rec{VirtualCircuitReq,\packcopy{}}$ originating in $n_0$ leads to the state $n_2$ witnessing the hazard $h \triangleq ( (\sigma_1,\sigma_2) ; (\neg VirtualCircuitEnd! \packcopy{} ) ^ *; (\sigma_3,\sigma_4)  ; A ^ * ) $. The next goal is to exploit the causal machinery in~\cite{DBLP:conf/tase/BonsangueCFT22} and derive
causal explanations for safety failures in SDN in an automated fashion.

\vspace{-10pt}
\section{Methodology, Results and Extensions}
The implementation for creating the LTS from DyNetKAT specifications is explained below and can be found at \url{https://github.com/canolmezoglu/DyNetiKAT}.

\smallskip\noindent
{\bf{Methodology:}}
To generate the LTS models, the prototype implementation \footnote{https://github.com/hcantunc/DyNetiKAT} from \cite{DBLP:conf/fossacs/CaltaisHMT22} was chosen as the base implementation for parsing an inputted DyNetKAT specification. This implementation was modified using Maude \cite{DBLP:conf/rta/ClavelDELMMQ99} to classify different operators of DyNetKAT. Following the parsing, we implemented in Python an algorithm that exploits the operational semantics from \cite{DBLP:conf/fossacs/CaltaisHMT22} and extracts the LTS from the  parsed specification. To obtain the causes from the LTS, we used Algorithm 1 from the work in \cite{DBLP:conf/tase/BonsangueCFT22}, where the LTS was converted into a FA model by considering all the states of the LTS as accepting states.

\smallskip\noindent
{\bf{Results:}}
 Upon conducting this methodology on the specification of the running example in Section \ref{sec:runningexample}, and using the regular expression $h$ as the hazard, four causal explanations can be identified as (the minimal) traces leading from $n_0$ to $n_2$.
 These traces entail a race condition arising between the controllers when the virtual circuit was first made active. 
 Note that, for the case study in this paper, there are no contingencies that can be used to steer the aforementioned causal explanations away from the undesired effect $h$. Hence, the actual causes coincide with the traces witnessing $h$ in the LTS model of the virtual circuit in~(\ref{eq:virtualcircuit}).

\smallskip\noindent
{\bf{Extensions:}}
Currently, we are working on developing a tool for extracting DyNetKAT specifications from real SDN data, based on the logs in \cite{DBLP:conf/pldi/El-HassanyMBVV16} and OpenFlow~\cite{DBLP:journals/ccr/McKeownABPPRST08}.
The latter is a protocol that can manipulate the control logic of a network and program the flow table of network switches. As OpenFlow networks are working, or when they are simulated, all the modifications are stored in the form of logs, describing what flow table updates have been made by which controller and for which network devices. Using these logs, such as ones that could be obtained from the work in \cite{DBLP:conf/pldi/El-HassanyMBVV16}, the state changes of the network switches can be inferred and converted into DyNetKAT specifications. These specifications can be used for causal analysis on real world data, as well as benchmarking the current prototype implementation.

\normalsize
\bibliographystyle{plain}



\end{document}